\documentclass[showpacs,showkeys,preprint,pra]{revtex4}
\usepackage{amsmath}

\begin{document}

\title{Reconsideration of quantum measurements}
\author{Spiridon Dumitru}
\email{s.dumitru@unitbv.ro} \affiliation{Departament of Physics,
"Transilvania" University, B-dul Eroilor 29, R-2200 Brasov,
Romania}

\begin{abstract}
The usual conjectures of quantum measurements approaches, inspired from the traditional interpretation of Heisenberg's 
("uncertainty") relations, are proved as being incorrect. A group of reconsidered conjectures and a corresponding new 
approach are set forth. The quantum measurements, regarded experimentally as statistical samplings, are described 
theoretically by means of linear integral transforms of quantum probability density and current, from  intrinsic into  
recorded readings. Accordingly, the quantum observables appear as random variables, valuable, in both readings,  
through probabilistic numerical parameters (characteristics). The measurements uncertainties (errors) are described by 
means of the intrinsic-recorded changes for the alluded parameters or for the informational entropies. The present 
approach, together with other author's investigations, give a natural and unified reconsideration of  primary problems 
of Heisenberg's relations and quantum measurements. The respective reconsideration can offer some nontrivial elements 
for an expected re-examination of some disputed subsequent questions regarding the foundations and interpretation of 
quantum mechanics.
\end{abstract}
\date{\today}
\maketitle

\section{Introduction}
In connection with the foundation and interpretation of quantum mechanics (QM) the problem of quantum measurements 
(QMS) description is of ten regarded \cite{1} as: "\emph{probably the most important part of the theory}". The 
respective problem consists in search for theoretical descriptions of the measurements regarding the observables of 
quantum microparticles. It was approached  in a large number of works published during the history of QM, particularly 
\cite{1} in the last decades. Many of the mentioned approaches were founded on conjectures, which someway are inspired 
from the early orthodox views on QM. Especially were referred the Heisenberg's ("uncertainly") relations (HR), 
currently regarded as corner stone for foundation and interpretation of QM. On the other hand nowadays \cite{2}: 
"\emph{the idea that there are defects in the foundation of orthodox quantum theory is unquestionable present in the 
conscience of many physicists}".

The spirit of the alluded idea motivated \cite{3,4,5,6} a careful  reinvestigation of the significance of HR. The 
reinvestigation  shows \cite{3,4,5,6} that HR must be reinterpreted in a more natural manner and, especially, deprived 
of all the attributes usually assumed by the QMS approaches. Then it directly appears the necessity of a 
reconsideration of QMS problem. Such a reconsideration is the aim of the present paper.

For our aim in the next section, we present the main alluded conjectures and their corresponding shortcomings. 
Subsequently in Sec.III we present a new approach of the QMS problem. The respective approach is inspired from a view 
\cite{7,8} about the measurements of classical (non-quantum) random observables. The new approach is detailed through 
a simple exemplification in Sec.IV. We end our considerations in Sec.V with some conclusions.

\section{Conjectures and shortcomings}
In the afferent publications, one finds a multitude of diverse approaches of QMS (for a comprehensive bibliography see 
\cite{1}). A careful examination of the things shows that, in their essence, many of the respective approaches imply 
someway one or more of the following conjectures(\textbf{C}):
\begin{itemize}
\item \textbf{C.1:} The description of QMS must be regarded and developed as an enlargement of the traditional 
interpretation of HR (TIHR).
\item \textbf{C.2:} Between quantum and classical measurements there exists a fundamental distinction, due to the 
exclusive existence of HR in quantum cases.
\item \textbf{C.3:} The description of QMS must take into account the non-null and unavoidable jumps in the state of 
the measured system.
\item \textbf{C.4:} A QMS consists in a single detection act representable as collapse (reduction) of  the wave 
function.
\end{itemize}
(Note that the conjectures \textbf{C.2-3} are inferable in part from \textbf{C.1} -see bellow).

The conjectures \textbf{C.1-3} are inspired from TIHR. Then, for a pertinent analysis of them, let us remind the main 
elements of TIHR. Therefore firstly we note that for two observables A and B the HR are known in the QM-theoretical 
version:
\begin{equation}\label{eq:1}
\Delta_{\Psi}A\cdot\Delta_{\Psi}B \geq \frac{1}{2}
\left\vert\langle[ \hat{A},\hat{B}]_{-}\rangle_{\Psi}\right\vert
\end{equation}
respectively  in the thought experimental (TE) version
\begin{equation}\label{eq:2}
\Delta_{TE}A\cdot\Delta_{TE}B \sim \hbar
\end{equation}
In (\ref{eq:1}) $[\hat{A},\hat{B}]_{-}=\hat{A}\hat{B}-\hat{B}\hat{A}$ denotes the commutator of the operators 
$\hat{A}$ and $\hat{B}$, while the other usual QM notations are those specified below in relations 
(\ref{eq:5})-(\ref{eq:6}). In (\ref{eq:2}) $\Delta_{TE}A$ and $\Delta_{TE}B$ signify the so-called TE-uncertainties 
for simultaneous measurement of two canonically conjugated (or complementary) observables $A$ and $B$ while $\hbar$ is 
the Planck's constant.

Based on the relations (\ref{eq:1})-(\ref{eq:2}) TIHR promoted the following main assertions (\textbf{Ass.}):
\begin{itemize}
\item \textbf{Ass.1:} The quantities $\Delta_{\Psi}A$ and $\Delta_{TE}A$ from (\ref{eq:1}) and (\ref{eq:2}) have the 
same significance of measuring uncertainty for $A$. Consequently, the respective relations refer to the description of 
a characteristic of QMS.
\item \textbf{Ass.2:} In QMS a single observable can be evaluated without any uncertainty while two observables $A$ 
and $B$ are compatible (i.e. simultaneously measurable without uncertainties) respectively incompatible as they are 
commutable or not (i.e. as $[\hat{A},\hat{B}]=0$ respectively $[\hat{A},\hat{B}]\neq 0$).
\item \textbf{Ass.3:} The relations (\ref{eq:1})-(\ref{eq:2}) do not have analogues in classical physics because they 
imply the quantum Planck's constant $\hbar$.
\end{itemize}

It is easy to see that the assertions \textbf{Ass.1-3} of THIR are irrefutably contradicted by the following arguments 
(\textbf{Arg.}):
\begin{itemize}
\item \textbf{Arg.1:} The relation (\ref{eq:1}) is in fact \cite{3,4,5,6} only a restricted consequence of the more 
general formula
\begin{equation}\label{eq:3}
\Delta_{\Psi}A\cdot\Delta_{\Psi}B\geq \left\vert (\delta_{\Psi}\hat{A}\Psi,\delta_{\Psi}\hat{B}\Psi)\right\vert
\end{equation}
(For the significance of the QM notations see below the relations (\ref{eq:5})-(\ref{eq:6})). Particularly there are 
situations in which (\ref{eq:1}) is inapplicable and only (\ref{eq:3}) remains valid in the trivial form $0=0$. Such 
is the case \cite{3,4,5,6} of pairs $L_{z}-\varphi$ (angular momentum-angle) and $N-\phi$ (number-phase) in 
eigenstates of $L_{z}$ respectively of $N$.
\item \textbf{Arg.2:} The observables implied in (\ref{eq:1}) and (\ref{eq:3}) have the characteristics of random 
variables. They are similar with the macroscopic observables from classical statistical physics \cite{9,10}. The 
respective similarity is pointed out by the existence \cite{11,12} of some fluctuation formulae which are completely 
analogous with (\ref{eq:1}) and (\ref{eq:3}). The alluded classical observables and formulae describe the macroscopic 
bodies themselves
(i.e. their intrinsic characteristic) but not the aspects of the measurements on such bodies. Similarly, relations 
(\ref{eq:1}) and (\ref{eq:3}) and the afferent quantities must be regarded as referring to the intrinsic properties 
(fluctuations) of quantum microparticles but not to the qualities (uncertainties) of QMS.
\item \textbf{Arg.3:} The relations (\ref{eq:2}) are founded only on the features of TE. However, as it is noted in 
\cite{13} there is not known any real experiment capable to attest TIHR with a convincing accuracy. On the other hand, 
the convincingness of the relations (\ref{eq:2}) is troubled by their provisional character. The respective character 
is related with the fact that the relations (\ref{eq:2}) were founded on old  classical limitative criteria (due 
\cite{14} to Abe and Rayleigh). However, in modern experimental physics \cite{15,16,17} super-resolution techniques 
which overstep the mentioned criteria are known. Then it is easy to see that (\ref{eq:2}) can be replaced by some 
super-resolution TE (SRTE) relations of the form
\begin{equation}\label{eq:4}
\Delta_{SRTE}A\cdot\Delta_{SRTE}B < \hbar
\end{equation}
where $\Delta_{SRTE}A$ and $\Delta_{SRTE}B$ denote the corresponding uncertainties.
Evidently the relations of type (\ref{eq:4}) incriminate the whole philosophy of TIHR. Now one can note that the above 
mentioned facts argue for the conclusion that relations (\ref{eq:2}) are only fictional formulas without any real 
value or significance for the description of QMS.
\item \textbf{Arg.4:} Conjointly with the facts noted in \textbf{Arg.1-3} one finds \cite{3,4,5,6} that the Planck's 
constant $\hbar$ does not play a role of finite lower bound for the products of measuring uncertainties. Moreover 
\cite{6,20,21} $\hbar$ proves oneself to be in the posture of a generic indicator of stochasticity (randomness) for 
quantum observables. In a similar posture, for classical macroscopic observables,there is the Boltzmann's constant $k$ 
\cite{12,20}.
\item \textbf{Arg.5:} The attunement of the energy-time pair with \textbf{Ass.2} (and consequently with TIHR doctrine) 
is in fact \cite{3,4,5,6} an impossible task. It generates only unproductive and interminable disputes. For the 
respective pair it must adopt a reasonable and natural regard. Such a regard can be obtained if, in quantum context, 
energy is considered as a random observable (endowed with possible fluctuations) while time is taken as a 
deterministic variable. Then for the energy time pair (\ref{eq:1}) is not applicable and only (\ref{eq:3}) remains 
valid in the trivial form $0=0$.
\item \textbf{Arg.6:} The assertion \textbf{Ass.2} irrevocably fails in some non-trivial situations regarding 
\cite{3,4,5,6} single observables respectively pairs of commutable observables.
\item \textbf{Arg.7:} It is notable \cite{3,4,5,6} the fact that \textbf{Ass.2} can not offer any reasonable base for 
the interpretation of some natural generalizations of the relations (\ref{eq:1}). Such is the case of bitemporal, 
many-observable and macroscopic-quantum-statistical generalizations.
\end{itemize}

The arguments \textbf{Arg.1-7} indubitably invalidate the whole class of assertions \textbf{Ass.1-3}. Consequently, 
TIHR must be denied as an unjustified doctrine. In addition, the HR (\ref{eq:1}) and (\ref{eq:2}) must be 
reinterpreted. In a genuine conception relation (\ref{eq:1}) appears \cite{3,4,5,6} as referring to intrinsic 
characteristics of quantum microparticles and belongs to a more large class of fluctuation formulas form both quantum 
and classical physics. In the same conception the relations (\ref{eq:2}) must be disregarded as fictitious formulas 
without any scientific significance. The mentioned denial of TIHR leaves without any base the conjectures 
\textbf{C.1-2}. Such a fact is an insurmountable shortcoming for all the QMS approaches, which imply the respective 
conjectures.

As regards to the conjectures \textbf{C.3} the following facts are notable. The respective conjecture was not inferred 
directly from the main assertions of TIHR. However, it was promoted adjacently in discussions generated by TIHR. 
Firstly, it was said that the measurements uncertainties are due to the interactions between measured systems and 
measuring devices. Secondly,  it was added that the respective interactions cause jumps in the states of the measured 
systems. Then it was accredited the supposition that, in contrast with the classical situations, in QMS the  mentioned 
uncertainties, interactions and jumps have an unavoidable character. Subsequently it was promoted the idea that the 
alluded measuring jumps must be taken into account in the description of QMS. In spite of its genesis, the above 
mentioned idea is proved to be incorrect by the following indubitable opinion \cite{22}: "\emph{it seems essential to 
the notion of a measurement that it answers a question about the given situation existing before measurement. Whether 
the measurement leaves the measured system unchanged or brings about a new and different state of that system is a 
second and independent question}".
The natural acceptance of the quoted opinion brings the conjecture \textbf{C.3} in an insurmountable shortcoming.

The conjecture \textbf{C.4} is contradicted by natural and authorized views from both physics and mathematics. From 
physical viewpoint, the measurement of a random observable must have the same general features, independently of its 
quantum or classical nature. However, in the classical context (e.g. in the study of fluctuation \cite{7,8,9,10,23}) 
the measurement of a macroscopic random observable is not viewed as a single detection act, associated with some 
collapse (reduction) of the corresponding probability distribution. More exactly such a measurement is regarded 
\cite{23} as a statistical sampling, i.e. as an ensemble of great number of individual detection acts. The respective 
ensemble gives a nontrivial set of values belonging to the spectrum of the considered observable. In addition, from a 
mathematical viewpoint \cite{24} a random variable must be evaluated not by a unique value but through a statistical 
set of values. Then it directly results that because QMS regards observables with random characteristics they must be 
viewed as statistical sampling (in the above-mentioned sense). Consequently there are no reason to represent 
(describe) a QMS as a collapse (reduction) of a wave function. The mentioned result and consequence incontestably 
invalidate the conjecture \textbf{C.4}. So one finds an insurmountable shortcoming for the respective conjecture.

The above presented aspects of the conjectures \textbf{C.1-4} show that in fact many of the proposed QMS approaches 
have important and unavoidable  shortcomings. Then it results that the problem of QMS description is still an open 
question that requires further investigations. The actuality of the alluded investigations is evidenced also by the 
nowadays scientific publications (see \cite{1} and references), gray literature \cite{25,26} respectively meetings 
\cite{27,28}. In such a context, we think that the new approach that we present in the next section can be of 
nontrivial interest.

\section{A new approach}
It is known that each approach of QMS description resorts (more or less explicitly) to some conjectures. Then, for the 
new approach aimed here, we suggest the set of the following reconsidered conjectures (\textbf{RC}):
\begin{itemize}
\item \textbf{RC.1} Any measurement search for information regarding the pre-existent state of the investigated 
system, independently of the quantum or classical nature of the respective system.
\item \textbf{RC.2} Due to the randomness of quantum systems a QMS must consists obligatory in a statistical sampling 
i.e. in a great number of individual detection acts.
\item \textbf{RC.3} QM refers to the intrinsic properties of the quantum systems and, consequently, a description of 
QMS must contain some extra-QM elements regarding the measuring devices and procedures.
\item \textbf{RC.4} Because, in the last analysis, the results supplied by QMS refer to the measured quantum systems 
they must be evaluated in terms of QM.
\end{itemize}

In mind with \textbf{RC.1-4}, we develop the announced approach as follows. We consider a spin-less quantum 
microparticle with own orbital characteristics described by the intrinsic (I) wave function $\Psi_{I}$. From 
theoretical viewpoint $\Psi_{I}$ can be regarded as solution of the corresponding Schr\" odinger equation. In the 
following probabilistic considerations, the microparticle is regarded as equivalent with a statistical ensemble of its 
own replica taken at the same instant of time and described by the same wave function $\Psi_{I}$. Therefore, for our 
purposes, the time $t$ appears as a "decorative" variable and $\Psi_{I}$ will be  written as a function only of the 
radius vector $\vec{r}$, i.e. $\Psi_{I}=\Psi_{I} (\vec{r})$. The specific observables $A_{j}\,(j=1,2,\ldots ,n)$ of  
the microparticle  are described by the usual QM operators (e.g. $\hat{x}_{\mu}=x_{\mu}\cdot$ and 
$\hat{p}_{\mu}=-i\hbar \frac{\partial}{\partial x_{\mu}}\,(\mu =1,2,3)$ for Cartesian coordinates and momenta, 
$\hat{\vec{p}}=-i\hbar \nabla$ and $\hat{\vec{L}}=-i\hbar \vec{r}\times \nabla$ for momentum and angular momentum 
vectors or $\hat{H}=-\frac{\hbar^{2}}{2m}\nabla^{2}+V\left(\vec{r}
\right)$ for Hamiltonian).

Because $A_{j}$ have random properties, as in probability theory \cite{24}, for practical purposes they are described 
by means of the so-called numerical parameters (or characteristics). In QM the mostly used such parameters are: the 
mean values $\langle A_{j} \rangle_{I}$, the correlations $\mathcal{C}_{I}(A_{j},A_{l})$, respectively the standard 
deviations $\Delta_{I}A_{j}$. Note that, from a probabilistic perspective, the mentioned numerical parameters are 
lower order entities. Additionally, as in probability theory \cite{24}, can be used also higher order numerical 
parameters (e.g. higher order correlations and moments). However, such parameters are not usual in QM literature. As 
it is known  the alluded lower order numerical parameters are defined by the relations:
\begin{equation}\label{eq:5}
\langle A_{j} \rangle_{I}=\langle\Psi_{I}\vert\hat{A}_j \Psi_{I}\rangle = \int \Psi_{I}^{*}\left( 
\vec{r}\right)\hat{A}_{j} \Psi_{I}\left( \vec{r} \right) \mathrm{d}^{3}\vec{r}
\end{equation}
\begin{equation}\label{eq:6}
\mathcal{C}_{I}(A_{j},A_{l})=\langle\delta_{I}\hat{A}_{j}\Psi{I} \vert \delta_{I}\hat{A}_{l}\Psi_{I}\rangle\;, \quad 
\delta_{I}\hat{A}_{j}=\hat{A}_{j}-\langle A_{j}\rangle_{I}
\end{equation}
\begin{equation}\label{eq:7}
\Delta_{I}A_{j}=\sqrt{\mathcal{C}_{I}(A_{j},A_{j})}
\end{equation}
In (\ref{eq:5}) and (\ref{eq:6}) $\langle f_{a}\vert f_{b}\rangle$ denotes the scalar product of the functions $f_{a}$ 
and $f_{b}$.
From measurements perspective the intrinsic parameters (\ref{eq:5})-(\ref{eq:7}) must be compared with the 
corresponding recorded (R) parameters:
\begin{equation}\label{eq:8}
\langle A_{j}\rangle_{R}\;, \qquad
\mathcal{C}_{R}(A_{j},A_{l})\;, \qquad
\Delta_{R}A_{j}
\end{equation}
In experimental approach the parameters (\ref{eq:8}) can be obtained by statistical processing of the data from the 
real measurements about the observables. On the other hand, if one wishes to operate with a description of the 
measurements, the parameters (\ref{eq:8}) must be regarded as pieces of an adequate theoretical model. For such a 
model we consider that the parameters (\ref{eq:8}) are defined similarly with (\ref{eq:5})-(\ref{eq:7}) by means of a 
"recorded"  wave function $\Psi_{R}$ and with the same QM operators, i.e.
\begin{equation}\label{eq:9}
\langle A_{j} \rangle_{R}=\langle\Psi_{R}\vert\hat{A}_j \Psi_{R}\rangle = \int \Psi_{R}^{*}\left( 
\vec{r}\right)\hat{A}_{j} \Psi_{R}\left( \vec{r} \right) \mathrm{d}^{3}\vec{r}
\end{equation}
\begin{equation}\label{eq:10}
\mathcal{C}_{R}(A_{j},A_{l})=\langle\delta_{R}\hat{A}_{j}\Psi{R} \vert \delta_{R}\hat{A}_{l}\Psi_{R}\rangle\;, \quad 
\delta_{R}\hat{A}_{j}=\hat{A}_{j}-\langle A_{j}\rangle_{R}
\end{equation}
\begin{equation}\label{eq:11}
\Delta_{R}A_{j}=\sqrt{\mathcal{C}_{R}(A_{j},A_{j})}
\end{equation}

     Our above consideration is motivated by the known fact that, in theoretical descriptions, the randomness of a 
quantum microparticle is incorporated in its wave function but not in operators of its observables. Properties of 
various states of a microparticle are described by different wave functions but with the same operators. A similar 
situation exists in the case of classical statistical systems for which the randomness is incorporated in the 
probability densities but not in the expressions of macroscopic random variables. In the alluded cases the properties 
of various states of a system are also described  with different probability densities but with the same expressions 
for the macroscopic random variables. In classical case a measurements described similarly \cite{7,8} by appealing to 
a "recorded"density of probability. Note  that in both quantum and classical cases the appeals to "recorded" entities 
(wave function or probability density) must not be regarded as a description of collapse (reduction) for the 
corresponding intrinsic entities.

By adopting the relations (\ref{eq:9})-(\ref{eq:11}) the task of our approach becomes to express $\Psi_{R}$(or related 
quantities) in terms of $\Psi_{I}$ (or associated  entities) and of some elements  regarding the measuring  devices. 
For such a task, firstly we show that the parameters (\ref{eq:5})-(\ref{eq:7}) and (\ref{eq:9})-(\ref{eq:11}) can be 
expressed in terms of certain quantities  connected with $\Psi_{Y}\,(Y=R,I)$ and having ordinary probabilistic 
significance in the probability theory sense \cite{24}. So we transcribe $\Psi_{Y}$in the form 
$\Psi_{Y}=|\Psi_{Y}|\exp (i \Phi_{Y})$, where $|\Psi_{Y}|$ and $\Phi_{Y}$ denote the modulus respectively the argument 
of $\Psi_{Y}$. As such ordinary quantities we take firstly the probability densities associated with $\Psi_{Y}$ and 
defined by
\begin{equation}\label{eq:12}
\rho_{Y}=|\Psi_{Y}|^{2}
\end{equation}
Other quantities with ordinary probabilistic significance are the probability currents (or probability fluxes per 
unit-area):
\begin{equation}\label{eq:13}
\vec{J}_{Y}=-\frac{i\hbar}{2m}\left( \Psi_{Y}^{*}\nabla\Psi_{Y} -\Psi_{Y}\nabla\Psi_{Y}^{*} 
\right)=\frac{\hbar}{m}|\Psi_{Y}|^{2} \cdot\nabla\Phi_{Y}
\end{equation}
($m$ denotes the mass of microparticle).

Now let us show that the parameters (\ref{eq:5})-(\ref{eq:7}) and (\ref{eq:9})-(\ref{eq:11}) can be expressed in terms 
of $\rho_{Y}$ and $\vec{J}_{Y}$. Then we observe that if an operator $\hat{A}$ does not depend on $\nabla$, i.e. 
$\hat{A}=\hat{A}(\vec{r})$ in (\ref{eq:5}) and (\ref{eq:9}) can be used the substitutions:
\begin{equation}\label{eq:14}
\Psi_{Y}^{*}\hat{A}\Psi_{Y}=A(\vec{r})\,\rho_{Y}
\end{equation}
On the other hand if $\hat{A}$ depends on $\nabla$, i.e. $\hat{A}=\hat{A}(\nabla)$, by taking $\Psi_{Y}=|\Psi_{Y}|\exp 
(i \Phi_{Y})$ and using (\ref{eq:12})-(\ref{eq:13}) in (\ref{eq:5}) and (\ref{eq:9}) one can resort to the 
substitutions like:
\begin{equation}\label{eq:15}
\Psi_{Y}^{*}\nabla\Psi_{Y}=\frac{1}{2}\nabla\rho_{Y}+ \frac{im}{\hbar}\vec{J}_{Y}
\end{equation}
\begin{equation}\label{eq:16}
\Psi_{Y}^{*}\nabla^{2}\Psi_{Y}=\rho_{Y}^{1/2}\nabla^{2}\rho_{Y}^{1/2} 
+\frac{im}{\hbar}\nabla\vec{J}_{Y}-\frac{m^2}{\hbar^2} \frac{\vec{J_Y}^2}{\rho_{Y}}
\end{equation}
The existence of substitutions (\ref{eq:14})-(\ref{eq:16}) suggests that the description of QMS can be completed by 
adequate considerations about the quantities $\rho_{Y}$ and $\vec{J}_{Y}$. As the respective quantities have ordinary 
probabilistic significance for the alluded completion we resort to the model used \cite{7,8} in the description of 
measurements of classical random observables. We also take into account the fact that $\rho_{Y}$ and $\vec{J}_{Y}$ 
refer to the positional respectively motional aspects of probabilities. Or, from an experimental perspective, the two 
aspects can be regarded as measurable by independent devices and procedures. Then the alluded completion must consists 
in giving independent relationships between $\rho_{R}$ and $\rho_{I}$ on the one hand respectively between 
$\vec{J}_{R}$ and $\vec{J}_{I}$ on the other hand.
The mentioned relationships can be expressed  formally by the following generic formulas:
\begin{equation}\label{eq:17}
\rho_{R}=\hat{G}\rho_{I}
\end{equation}
\begin{equation}\label{eq:18}
J_{R;\mu}=\sum_{\nu=1}^{3}\hat{\Lambda}_{\mu\nu}J_{I;\nu}
\end{equation}
($J_{Y;\mu}$ with $Y=R,I$ and $\mu=1,2,3=x,y,z$ denote the
Cartesian components of  vectors $\vec{J}_{Y}$). In (\ref{eq:17}) and (\ref{eq:18}) $\hat{G}$ and 
$\hat{\Lambda}_{\mu,\nu}$ signify the measurements operators. They must comprise obligatory characteristics of 
measuring devices and procedures. So $\hat{G}$ and $\hat{\Lambda}_{\mu,\nu}$ must contain some extra-QM elements, i.e. 
elements that do not belong to the usual QM description of the intrinsic properties of the measured microparticles.

For measuring devices with linear and stationary characteristics, similarly with the classical case \cite{7,8}, the 
relations (\ref{eq:17})-(\ref{eq:18}) can be written as:
\begin{equation}\label{eq:19}
\rho_{R}(\vec{r})=\int G(\vec{r},\vec{r}\,')\, \rho_{I}( \vec{r}\,')\,\mathrm{d}^{3}\vec{r}\,'
\end{equation}
\begin{equation}\label{eq:20}
J_{R;\mu}(\vec{r})=\sum_{\nu=1}^{3}\int \Lambda_{\mu\nu} 
(\vec{r},\vec{r}\,')\,J_{I;\nu}(\vec{r}\,')\,\mathrm{d}^{3}\vec{r}\,'
\end{equation}
The kernels $G(\vec{r},\vec{r}\,')$ and $\Lambda_{\mu\nu}(\vec{r},\vec{r}\,')$ are supposed to satisfy the conditions:
\begin{equation}\label{eq:21}
\int G(\vec{r},\vec{r}\,')\,\mathrm{d}^{3}\vec{r}= \int G(\vec{r},\vec{r}\,')\, \mathrm{d}^{3}\vec{r}\,'=1
\end{equation}
\begin{equation}\label{eq:22}
\int \Lambda_{\mu\nu}(\vec{r},\vec{r}\,')\,\text{d}^{3}\vec{r} = \int 
\Lambda_{\mu\nu}(\vec{r},\vec{r}\,')\,\text{d}^{3}\vec{r}\,'=1
\end{equation}
These conditions show the one-to-one probabilistic correspondence between the intrinsic quantities $\rho_{I}$ and 
$\vec{J}_{I}$ respectively the recorded ones $\rho_{R}$ and $\vec{J}_{R}$.
Parameters (\ref{eq:9})-(\ref{eq:11}), evaluated by means of the relations (\ref{eq:14})-(\ref{eq:16})and 
(\ref{eq:17})-(\ref{eq:22}), incorporate randomness of both intrinsic and extrinsic nature, corresponding to the own 
properties of the investigated microparticle respectively to the measuring devices. As evaluated the mentioned 
parameters have a theoretical significance. Their adequacy must be tested by comparing with the corresponding 
parameters (\ref{eq:8}) obtained by statistical processing of the real experimental data. If the test is affirmative 
both descriptions, of intrinsic QM properties respectively of QMS, can be accepted as adequate. However, if the test 
invalidates the theoretical results, at least one of the respective descriptions must be regarded as  inadequate.

From the origins of their history, the QMS approaches are concerned with the problem of quantitative evaluation for 
measuring uncertainties (i.e. for errors induced by the measurements in the values of the measured quantum 
observables). That is why it is of interest to discuss the respective problem in connection with the here promoted 
approach. Our discussion starts by pointing out the fact that quantum observables have a random character. 
Consequently, the uncertainties of such an observable must be evaluated through   indicators, which comprise 
information from the whole its spectrum. It is easy to see that indicators of the alluded kind can be introduced by 
means of the numerical parameters defined by relations (\ref{eq:5})-(\ref{eq:7}) and (\ref{eq:9})-(\ref{eq:11}). That 
is why we suggest that, conjointly with the above--presented approach of QMS, the measuring uncertainties to be 
evaluated through the following     error (or uncertainty) indicators:
\begin{equation}\label{eq:23}
\delta\left( \langle A_{j} \rangle \right)=\left| \langle A_{j} \rangle _{R} - \langle A_{j} \rangle _{I} \right|
\end{equation}
\begin{equation}\label{eq:24}
\delta \left( \mathcal{C}(A_{j},A_{l}) \right)= \left| \mathcal{C}_{R}(A_{j},A_{l})-\mathcal{C}_{I}(A_{j},A_{l}) 
\right|
\end{equation}
\begin{equation}\label{eq:25}
\delta (\Delta A_{j})=\left| \Delta_{R}A_{j}-\Delta_{I}A_{j} \right|
\end{equation}
These indicators have a restricted significance for a system (microparticle), because they refer to some particular 
observables of the respective system. A more generic uncertainty indicators, regarding a system in the whole, can be 
introduced by means of informational entropies.

In connection with the applicability of the informational entropies for quantum cases a recent opinion is known 
\cite{29}. According to the respective opinion the Shanonn entropy is inadequate for the respective cases because of  
the contrast between the quantum and classical situations. The contrast is associated with ideas such are:  (i) QMS 
\cite{29}: "\emph{cannot be claimed to reveal a property of the individual quantum system existing before the 
measurement is performed}" or (ii) The description of QMS must be considered joint with the quantum complementarity.

One can see that the mentioned ideas are reminiscences of the QMS approaches based on TIHR. However, as we have argued 
above such approaches are incorrect. On the other hand the quantities $\rho_{Y}$ and $\vec{J}_{Y}$ used in our 
approach have all the essential characteristics of classical probabilities. Therefore, for our purposes, we can 
operate with entropies of Shannon type. Note that for other purposes the  informational entities (different from 
Shannon entropy) proposed or reminded in \cite{29} can be of real utility. Here we use the following informational 
entropies of Shannon type:
\begin{equation}\label{eq:26}
\mathcal{H}_{Y}=-\int \rho_{Y} \ln \rho_{Y}\, \mathrm{d}^{3}\vec{r}
\end{equation}
\begin{equation}\label{eq:27}
\tau_{Y}=-\int |\vec{J}_{Y}| \ln |\vec{J}_{Y}|\, \mathrm{d}^{3}\vec{r}
\end{equation}
Here $\mathcal{H}_{Y}$ and $\tau_{Y}$ can be called positional respectively motional informational entropies. Then the 
alluded generic uncertainty indicators can be defined as
\begin{equation}\label{eq:28}
\delta \mathcal{H}=\mathcal{H}_{R}-\mathcal{H}_{I}
\end{equation}
\begin{equation}\label{eq:29}
\delta\tau=\tau_{R}-\tau_{I}
\end{equation}
It is interesting to note the fact that within the above-presented description of QMS the indicator 
$\delta\mathcal{H}$ is a nonnegative quantity (i.e. $\delta\mathcal{H}\geq 0$).
The respective fact can be proved, similarly with the classical situation \cite{7,8}, by means of the relations 
(\ref{eq:19}) and (\ref{eq:21}). So by taking into account the respective relations, the normalization of both 
$\rho_{I}$ and $\rho_{R}$, and the evident formula $\ln y \leq y-1\; (y>0)$ one can write:
\begin{eqnarray}\label{eq:30}
\delta\mathcal{H}&=&\mathcal{H}_{R}-\mathcal{H}_{I}=\nonumber \\
&=&-\int \mathrm{d}^{3}\vec{r} \int \mathrm{d}^{3}\vec{r}\,'\, G(\vec{r},\vec{r}\,')\, \rho_{I}(\vec{r}\,')^{2} \ln 
\frac{\rho_{R}(\vec{r})}{\rho_{I}(\vec{r}\,')} \geq \nonumber \\
&\geq& -\int \mathrm{d}^{3}\vec{r} \int \mathrm{d}^{3}\vec{r}\,'\,G(\vec{r},\vec{r}\,')\, \rho_{I}(\vec{r}\,') \left[ 
\frac{\rho_{R}(\vec{r})}{\rho_{I}(\vec{r}\,')} -1 \right]=0
\end{eqnarray}

 The above considerations give a natural description of QMS in which one finds, in adequate positions, all the 
essential elements. The respective elements include: (i) the intrinsic numerical parameters (\ref{eq:5})-(\ref{eq:7}), 
(ii) the model represented by (\ref{eq:17})-(\ref{eq:22}) for describing the influences of measuring devices, (iii) 
the recorded numerical parameters (\ref{eq:9})-(\ref{eq:11}) and (iv) the uncertainties indicators 
(\ref{eq:23})-(\ref{eq:25}) or (\ref{eq:28})-(\ref{eq:29}).

In the end of this section we note  that the description of QMS presented here, as well as the one discussed in 
\cite{7,8} for classical measurements, can be regarded formally from the perspective of information theory. In such a 
perspective, a measurement appears as a process of information transmission. The source of information is the measured 
system and the intrinsic values of its random characteristics (probability density and current, or observables) 
represent the input information. The chain of measuring devices play the role of channel for information transmission. 
The recorded data about the measured random characteristics represent the output information. Then the measurement 
uncertainties can be regarded as alterations of the transmitted information.

\section{A simple exemplification}
To illustrate the above-introduced QMS approach let us refer to the following simple exemplification. We consider a 
quantum  microparticle in a one-dimensional motion along the $x$-axis. Its own properties are supposed to be described 
by the intrinsic wave function $\Psi_{I}(x) =|\Psi_{I}(x)|\exp\left( i\Phi_{I}(x) \right)$ with:
\begin{equation}\label{eq:31}
\Psi_{I}(x)=\left( \alpha \sqrt{2\pi} \right)^{-\frac{1}{2}} \exp\left\{ -\frac{(x-x_{0})^{2}}{4\alpha^{2}} 
\right\}\,,
\quad \Phi(x)=kx
\end{equation}
Then the intrinsic probability density and current defined by (\ref{eq:12}) and (\ref{eq:13}) are:
\begin{equation}\label{eq:32}
\rho_{I}(x)=\frac{1}{\alpha\sqrt{2\pi}}\, \exp{\left\{{-\frac{(x-x_{0})^{2}}{2\alpha^{2}}}\right\}}
\end{equation}
respectively
\begin{equation}\label{eq:33}
J_{I}(x)=\frac{\hbar k}{m\alpha\sqrt{2\pi}}
\exp{\left\{{-\frac{(x-x_{0})^{2}}{2\alpha^{2}}}\right\}}
\end{equation}
So the intrinsic characteristics of the microparticle are described by the parameters $x_{0}$, $\alpha$ and $k$.

Considering that the errors of QMS are small in (\ref{eq:19}) and (\ref{eq:20}), one can operate with the 
one-dimensional kernels of Gaussian forms given by:
\begin{equation}\label{eq:34}
G(x,x')=\frac{1}{\sigma \sqrt{2\pi}}
\exp{\left\{ -\frac{(x-x')^{2}}{2\sigma^{2}} \right\}}
\end{equation}
\begin{equation}\label{eq:35}
\Lambda(x,x')=\frac{1}{\lambda\sqrt{2\pi}}
\exp{\left\{ -\frac{(x-x')^{2}}{2\lambda^{2}} \right\}}
\end{equation}
Here $\sigma$ and $\lambda$ describe the error characteristics of the measuring devices (see bellow).

By using (\ref{eq:34})-(\ref{eq:35}) in the one-dimensional versions of the relations (\ref{eq:19})-(\ref{eq:20}) one 
finds:
\begin{equation}\label{eq:36}
\rho_{R}(x)= \frac{1}{\sqrt{2\pi(\alpha^{2}+\sigma^{2})}}
\exp{\left\{ -\frac{(x-x_{0})^{2}}{2(\alpha^{2}+\sigma^{2})} \right\}}
\end{equation}
\begin{equation}\label{eq:37}
J_{R}(x)=\frac{\hbar k}{m\sqrt{2\pi(\alpha^{2}+\lambda^{2})}}
\exp{\left\{ -\frac{(x-x_{0})^{2}}{2(\alpha^{2}+\lambda^{2})} \right\}}
\end{equation}
One can see that in the case when $\sigma\to 0$ and $\lambda \to 0$ the kernels $G(x,x')$ and $\Lambda(x,x')$ 
degenerate into the Dirac function $\delta(x-x')$. Then $\rho_{R}(x)\rightarrow \rho_{I}(x)$ and $J_{R}(x)\rightarrow 
J_{I}(x)$. Such a case corresponds to an ideal measurement. Alternatively the cases with $\sigma\neq 0$and/or 
$\lambda\neq 0$ are associated with non-ideal measurements.

As observables of interest, we consider the coordinate $x$ and momentum $p$ described by the operators 
$\hat{x}=x\cdot$ and $\hat{p}= -i\hbar\frac{\partial}{\partial x}$. Adequately we use the expressions 
(\ref{eq:32})-(\ref{eq:33}) and (\ref{eq:36})-(\ref{eq:38}) in the relations (\ref{eq:5})-(\ref{eq:6}) and 
(\ref{eq:9})-(\ref{eq:11}).
Then, by using (\ref{eq:14})-(\ref{eq:16}), for the mentioned obsevables one finds the following intrinsic (I) 
respectively recorded (R) numerical parameters:
\begin{equation}\label{eq:38}
\langle x \rangle_{I}= \langle x \rangle_{R}=x_{0}\;,  \qquad
\langle p \rangle_{I}= \langle p \rangle_{R}=\hbar k
\end{equation}
\begin{equation}\label{eq:39}
\mathcal{C}_{I}(x,p)=\mathcal{C}_{R}(x,p)=\frac{i\hbar}{2}
\end{equation}
\begin{equation}\label{eq:40}
\Delta_{I}x=\alpha\;, \qquad
\Delta_{R}x=\sqrt{\alpha^{2}+\sigma^{2}}
\end{equation}
\begin{equation}\label{eq:41}
\Delta_{I}p=\frac{\hbar}{2\alpha}
\end{equation}
\begin{equation}\label{eq:42}
\Delta_{R}p=\hbar \sqrt{\frac{k^{2}(\alpha^{2}+\sigma^{2})} 
{\sqrt{\alpha^{4}-\lambda^{4}+2\sigma^{2}(\alpha^{2}+\lambda^{2})}} -k^{2}+\frac{1}{4(\alpha^{2}+\sigma^{2})}}
\end{equation}
Then for the considered observables $x$ and $p$ the error indicators (\ref{eq:23})-(\ref{eq:25}) become:
\begin{equation}\label{eq:43}
\delta\left( \langle x \rangle \right)=0 \,, \quad
\delta\left( \langle p \rangle \right)=0 \,, \quad
\delta\left( \mathcal{C} (x,p) \right)=0
\end{equation}
\begin{equation}\label{eq:44}
\delta \left( \Delta x \right) = \sqrt{\alpha^{2}+\sigma^{2}}-\alpha
\end{equation}
\begin{equation}\label{eq:45}
\delta\left( \Delta p \right)=\hbar \left\{
\sqrt{\frac{k^{2}(\alpha^{2}+\sigma^{2})} {\sqrt{\alpha^{4}-\lambda^{4}+2\sigma^{2}(\alpha^{2}+\lambda^{2})}} 
-k^{2}+\frac{1}{4(\alpha^{2}+\sigma^{2})}} -\frac{1}{2\alpha}\right\}
\end{equation}

These relations show that for the considered association microparticle-QMS the numerical parameters $\langle 
x\rangle$, $\langle p\rangle$ and $\mathcal{C}(x,p)$ are not affected  by errors. However, for the same  association 
the parameters $\Delta x$ and $\Delta p$ are troubled by the measurement, the corresponding non-null error indicators 
being given by (\ref{eq:44})-(\ref{eq:45}). Then, within the discussed model for the pair of non-commutable 
observables $x$ and $p$ one finds the relations:
\begin{equation}\label{eq:46}
\delta\left( \langle x \rangle \right)\cdot \delta\left( \langle p \rangle \right)=0
\end{equation}
\begin{equation}\label{eq:47}
\delta \left( \Delta x \right)\cdot \delta \left( \delta p \right) =\hbar \varepsilon
\end{equation}
where $\varepsilon$ can be identified from (\ref{eq:44})-(\ref{eq:45}).

Relations (\ref{eq:46})-(\ref{eq:47}) can be offered as pieces for a natural reconsideration of the above-mentioned 
assertion \textbf{Ass.2} of TIHR regarding the non-commutable observables. Of course that the offer is accompanied by 
the observation that, as can be see from (\ref{eq:44}),(\ref{eq:45}) and (\ref{eq:47}), $\varepsilon\to 0$ when 
$\sigma\to 0$ and $\lambda\to 0$ (i.e. for ideal measurements). Then it results that on principle the products 
$\delta(\langle x\rangle)\cdot \delta(\langle p\rangle)$ and $\delta(\Delta x)\cdot \delta(\Delta p)$ does not have 
non-null lower bounds. But such a fact contradicts the essential idea of TIHR that for  non-commutable  observables 
the measuring uncertainties (errors) are mutually lower bounded, independently of the concrete performances of the 
experiments.

Now, for the here discussed model of QMS description, let us  search the entropic error indicators defined by the 
relations (\ref{eq:26})-(\ref{eq:29}). By using the expressions (\ref{eq:32})-(\ref{eq:36}) one finds:
\begin{equation}\label{eq:48}
\delta\mathcal{H}=\frac{1}{2}\ln{\left( 1+\frac{\sigma^{2}}{\alpha^{2}} \right)}
\end{equation}
\begin{equation}\label{eq:49}
\delta\tau =\frac{\hbar k}{2m}\ln{\left( 1+ \frac{\lambda^{2}}{\alpha^{2}}\right)}
\end{equation}
If in (\ref{eq:31}) we choose $x_{0}=0$, $k=0$ and $\alpha=\sqrt{\frac{\hbar}{2m\omega}}$, our system is just a 
quantum  oscillator with mass $m$ and pulsation $\omega$ situated in its ground state. The corresponding statistical 
estimators and error indicators for observables $x$ and $p$ can be obtained from (\ref{eq:39})-(\ref{eq:43}) 
respectively (\ref{eq:44})-(\ref{eq:46}) by mentioned choosing. However, in the case of oscillator it is interesting 
to point out the measuring characteristics for another observable, described by the Hamiltonian:
\begin{equation}\label{eq:50}
\hat{H}=\frac{1}{2m}\hat{p}^2+\frac{m\omega^2}{2}\hat{x}^2
\end{equation}
Then, by using (\ref{eq:18})-(\ref{eq:20}) and (\ref{eq:33}), for the numerical parameters of oscillator energy one 
finds:
\begin{equation}\label{eq:51}
\langle H\rangle_{I}=\frac{\hbar\omega}{2}\;, \qquad
\Delta_{I}H=0
\end{equation}
\begin{equation}\label{eq:52}
\langle H\rangle_{R}=\frac{\omega \left[ \hbar^2+\left(\hbar+2m\omega \sigma^2\right)^2 
\right]}{4(\hbar+2m\omega\sigma^2)}
\end{equation}
\begin{equation}\label{eq:53}
\Delta_{R}H=\frac{2m\omega^{2}\sigma^{2}(\hbar+m\omega\sigma^{2})}{\sqrt{2}(\hbar+2m\omega\sigma^{2})}
\end{equation}
The corresponding uncertainty (error) indicators are:
\begin{equation}\label{eq:54}
\delta\left(\langle H\rangle\right)=\frac{\omega \left[ \hbar^2+\left(\hbar+2m\omega \sigma^2\right)^2 
\right]}{4(\hbar+2m\omega\sigma^2)}-\frac{\hbar\omega}{2}
\end{equation}
\begin{equation}\label{eq:55}
\delta(\Delta H)=\frac{2m\omega^{2}\sigma^{2}(\hbar+m\omega\sigma^{2})}{\sqrt{2}(\hbar+2m\omega\sigma^{2})}
\end{equation}

\section{Conclusions}
The problem of QMS description persists in our days as an open and disputed question. Many of its approaches are 
founded on fictitious conjectures, mainly inspired from TIHR. We aimed  here a new approach started with an 
investigation of the correctness of the alluded conjectures. So we found in fact a true incorrectness, the respective 
conjectures being affected by insurmountable shortcomings. Such a finding motivates our interest for a new set of 
reconsidered and natural conjectures. Consequently, we proposed a set of four such conjectures. Guided by that 
proposal we developed a new approach for the description of QMS.

Our approach is founded on the usual probabilistic conception of QM.  Therefore, for a quantum microparticle we 
operate with probability density, probability current and QM operators. We opine that, because in practice a correct 
QMS must consist in a statistical sampling, from a theoretical viewpoint a QMS must be represented as processing of 
the mentioned probabilistic entities. Similarly, with the description of classical (non-quantum) measurements, the 
alluded processing must be pictured as changes of the respective entities. We opine that for a wide class of 
situations such changes can be modelled  as linear integral transforms. Therefore, both probability density and 
probability current appear in intrinsic respectively recorded posture. In the first posture, they regard the own 
characteristics of the measured micropartice, while in the second posture they comprise information related both with 
the respective microparticle and with the measuring devices.

Together with the mentioned features of QMS the quantum observables must be naturally evaluated through the numerical 
parameters such are: mean values, correlations and standard deviations. Within the discussed approach, the respective 
parameters  are characterized by intrinsic respectively recorded values. Then a natural description of measuring 
uncertainties for quantum observables is expressible in terms of differences between the mentioned recorded and 
intrinsic values. Another description of measurements uncertainties, more generic (i.e. not associated with some 
particular observables), can be done in terms of informational entropies.

The here recapitulated features of our QMS approach are detailed from a general perspective in Section III, while in 
Section IV they are illustrated by means of a simple exemplification.

We remind here that our QMS approach is quite different from the approaches founded on (or inspired from) TIHR. The 
difference is evidenced on the one hand by the idea that QMS must be regarded as statistical samplings but not as 
individual detection acts. Consequently, we can avoid completely the controversial conception of wave function 
collapse (reduction). On the other hand, the alluded difference is pointed out by our natural presumtion that the 
description of QMS must be regarded as distinct and independent task comparatively with the QM investigations. 
Accordingly, with the respective regards the description of measurement must be considered and discussed as a 
scientific branch self-determined and additional comparatively with the quantum or classical chapters of physics. The 
mentioned chapters, as in fact is well established by the scientific practice, investigate only the intrinsic 
properties of the physical systems.

The views promoted here and in \cite{3,4,5,6} give a natural and unified reconsideration of the prime problems 
regarding HR and QMS. The problems refer to the interpretation of HR respectively to the description of QMS and the 
reconsideration is founded on an argued denial of TIHR. On the other hand HR and QMS are also implied in many 
subsequent questions. The questions regard the foundations and interpretation of QM and have been largely disputed in 
the last years (for an actualized bibliography see the works \cite{1,6,30,31,32} with their references as well as the 
gray literature sources \cite{25,26}).  In the respective questions HR and QMS are taken as pre-existing entities with 
properties inferred from various presumptions. Surprisingly, in many disputed cases the alluded presumptions originate 
(more or less explicitly) from TIHR. Therefore, due to the mentioned denial of TIHR, it is thinkable that our 
reconsideration of HR and QMS can offer nontrivial elements for an expected re-examination of some from the alluded 
subsequent questions. Of course that the respective elements can (and must) be complemented with other considerations 
regarding the QM problems.

\begin{acknowledgments}
I wish to express my deep gratitude to the World Scientific Publishing Company for putting at my disposal a copy of 
the monumental book \cite{1}.

The investigations reported in the present text benefited partially from some facilities of a grant from the Romanian 
Ministry of Education and Research.
\end{acknowledgments}


\begin{references}
\bibitem{1} G. Auletta, \textit{Foundations and Interpretation of Quantum Mechanics} (World Scientific, Singapore 
2000).
\bibitem{2} C. Piron, Lect. Notes. Phys \textbf{153}, 179 (1982).
\bibitem{3} S. Dumitru, Epistemological Lett. \textbf{15}, 1 (1977).
\bibitem{4} S. Dumitru, in \textit{Recent Advances in Statistical Physics}, Proceedings of International Bose 
Symposium on Statistical Physics, Calcutta, India, 1984, edited by B.Datta and M. Dutta (World Scientific, Singapore 
1987) p.122-151.
\bibitem{5} S. Dumitru, Rev. Roum. Phys. \textbf{33}, 11 (1988).
\bibitem{6} S. Dumitru, e-print: http://xxxaps.org/quant-ph/0004013
\bibitem{7} S. Dumitru, Phys. Lett. \textbf{A48}, 109 (1974).
\bibitem{8} S. Dumitru, Optik (Stuttgart) \textbf{110}, 110 (1999).
\bibitem{9} L. D. Landau, E. M. Lifchitz, \textit{Physique Statistique} (Mir, Moscou 1984).
\bibitem{10} B. Diu, C. Guthmann, D. Lederer, and B. Roulet, \textit{Elements de Physique Statistique} (Hermann, Paris 
1995).
\bibitem{11} S. Dumitru, Phys.Scr. \textbf{10}, 101 (1974).
\bibitem{12} S. Dumitru, A. Boer, Phys. Rev. E \textbf{64}, 021108 (2001).
\bibitem{13} M. Jammer, \textit{The Philosophy of Quantum Mechanics} (J.Wiley, New York 1974).
\bibitem{14} M. Born, E. Wolf, \textit{Principles of Optics} (Pergamon  Press, Oxford 1968).
\bibitem{15} C. Rychoudhuri, Found.Phys. \textbf{8}, 845 (1978).
\bibitem{16} J. Scheer, T. G\"{o}tsch, T. Coch, G. L\"{u}ning, M. Schmidt and H. Ziggel, Found.Phys.Lett. \textbf{2}, 
71 (1989).
\bibitem{17} J. R. Croca, A.Rica da Silva and J.S.Ramos, "\textit{Experimental Violation of Heisenberg's Uncertainty 
Relations by the Scanning Near-Field Optical Microscope}" Preprint (University of Lisbon, Portugal 1996). (This work 
discusses the potential implications of performances attained in optical experiments such are those reported in 
\cite{18,19}.
\bibitem{18} D. W. Pohl, W. Denk and M. Lanz, Appl. Phys. Lett. \textbf{44}, 651 (1994).
\bibitem{19} H. Heiselman and D. W. Pohl, Appl. Phys. A, \textbf{58}, 89 (1994).
\bibitem{20} S. Dumitru, Physics Essays \textbf{6}, 5 (1993).
\bibitem{21} S. Dumitru, E. I. Verriest, Int. J. Theor. Phys. \textbf{34}, 1785 (1995).
\bibitem{22} J. Albertson, Phys. Rev. \textbf{129}, 940 (1963).
\bibitem{23} H. Shilling, \textit{Statistiche Physik in Beispielen} (Veb Fachbucherverlag, Leipzig, l972).
\bibitem{24} G. A. Korn, T. M. Korn, \textit{Mathematical Handbook} (Mc Graw Hill, New York 1968) (Russian version: 
Nauka, Moscow 1977).
\bibitem{25} Lists \textit{New Preprints and Reports in the CERN Library} http://weblib.cern.ch
\bibitem{26} arXiv.org e-Print archive - Quantum Physics: http://mentor.lanl.gov/archive/quant-ph
\bibitem{27} \textit{Foundations of Probability and Physics} - \mbox{International Conference: 
2000 Nov. 27} - Dec. 1, Vaxjo University, Sweden http://www.msi.vxu.se/aktuellt/konferens/foundations.html
\bibitem{28} \mbox{\textit{Quantum Theory: Reconsideration of Foundations} - International Conference:} 2001 June 
17-21, Vaxjo University, Sweden. http://www.msi.vxu.se/aktuaellt/konferens/quantumtheory.html
\bibitem{29} C. Brukner, A. Zeilinger, Phys. Rev. A \textbf{63}, 022113 (2001).
\bibitem{30} M.B. Menskii, Physics-Uspekhi \textbf{43}, 585 (2000).
\bibitem{31} \textit{Letters to the Editor}, Physics-Uspekhi \textbf{44}, p.417-442 (2001).
\bibitem{32} F. Laloe, Am.J.Phys. \textbf{69}, 655 (2001).
\end{references}
\end{document}